\documentclass[fleqn,12pt]{article}
\usepackage{oldlfont,alltt,multicol,epsfig,amsmath}
\newcommand{\mydef}{\stackrel{\hbox{\scriptsize\textrm{def}}}{=}}
\newcommand{\vol}{\hbox{\textrm{vol}}}

\title{Efficient cache use for stencil operations on structured
       discretization grids}
\author{Michael A.\ Frumkin and Rob F.\ Van der Wijngaart\thanks{Computer
        Sciences Corporation; M/S T27A-2, NASA Ames Research Center,
        Moffett Field, CA 94035-1000; e-mail:
        \texttt{\{frumkin,wijngaar\}@nas.nasa.gov}}\\
        Numerical Aerospace Simulation Systems Division \\
        NASA Ames Research Center}
\date{}
\begin{document}
\maketitle
\begin{abstract}
We derive tight bounds on the cache misses for evaluation of explicit stencil
operators on structured grids.
Our lower bound is based on the isoperimetrical property of the discrete
octahedron.
Our upper bound is based on a good surface to volume ratio of a parallelepiped
spanned by a reduced basis of the interference lattice of a grid.
Measurements show that our algorithm typically reduces the number of cache 
misses by a factor of three, relative to a compiler optimized code.
We show that stencil calculations on grids whose interference lattice
have a short vector feature abnormally high numbers of cache misses.
We call such grids unfavorable and suggest to avoid these in computations
by appropriate padding.
By direct measurements on a MIPS R10000 processor we show a good correlation
between abnormally high numbers of cache misses and unfavorable 
three-dimensional grids.
\end{abstract}
\section{Introduction}
On modern computers the gap between access times to cache and to global
memory amounts to several orders of magnitude, and is growing.
As a result, improvement in usage of the memory hierarchy has become a
significant source of enhancing application performance.
Well-organized data traffic may improve performance of a program,
without changing the actual amount of computation, by reducing the time
the processor stalls waiting for data.
Both data location and access patterns affect the amount of data 
movement in the program, and the effectiveness of the cache.

A number of techniques for improvements in usage of data caches have
been developed in recent years.
The techniques include improvements in data reuse (i.e.\ temporal
locality) \cite{COLEMAN,MARTONOSI,MARTONOSI2,WOLF}, improvements in data locality 
(i.e.\ spatial locality) \cite{WOLF}, and reductions in conflicts in data 
accesses \cite{MARTONOSI,MARTONOSI2,RIVERA1,RIVERA2}.
In practice, these techniques are implemented through code and data
transformations such as array padding and loop unrolling, tiling, and fusing.
Tight lower and upper bounds on memory hierarchy access complexity for
FFT and matrix multiplication algorithms are given in \cite{PEBBLE}.
However, questions concerning bounds on the number of cache misses and
how closely current optimization techniques approach those bounds for stencil
operators remain open.

In this paper we consider improvement of cache usgae through maximizing 
temporal locality in evaluations of explicit stencil operators on
structured discretization grids.
Our contribution if twofold.
First, we prove lower and upper bounds on the number of cache misses for
local operators on structured grids.
Our lower bound (i.e.\ the number of unavoidable misses) is based on the
discrete isoperimetric theorem.
Our upper bound (i.e.\ the achievable number of misses) is based on a cache 
fitting algorithm which utilizes a special basis of the grid interference
lattice.
As shown by example, the lower bound can be achieved in some cases.
The second contribution is the identification of grids unfavorable
dimensions which cause significant increases in cache misses.
We provide two characterizations of these unfavorable grids.
The first one, derived experimentally, states that the product of
all relevant grid dimensions is close to a multiple of half the cache size.
The second characterization is that the grid interference lattice has a short
vector.
\section{Cache model and definitions}
We consider a single-level, virtual-address-mapped, set-associative data cache
memory, see \cite{HENNESSY}.
The memory is organized in $a$ sets of $z$ lines of $w$ words each.
Hence, it can be characterized by the parameter triplet $(a,z,w)$, and its
size $S$ equals $a*z*w$ words.
A cache with parameters $(a,1,w)$ is called fully associative, and with 
parameters $(1,z,w)$ it is called direct-mapped.

The cache memory is used as a temporary fast storage of words used for
processing.
A word at virtual address $A$ is fetched into a $(a(A),z(A),w(A))$ cache
location, where $w(A) = A \bmod w$, $z(A) = (a/w) \bmod z$, and $a(A)$
is determined according to a replacement policy (usually a variation of
\textit{least recently used}).
The replacement policy is not important within the scope of this paper.

If a word is fetched, then $w-1$ neighboring words are fetched as well
to fill the cache line completely.
In practice, $a$, $z$, and $w$ are often powers of 2 in order to
simplify computation of the location in cache.
For example, the MIPS R10000 processor for which we report some
measurements in Section \ref{sec:RESULTS}, has a cache with parameters
(2,512,4), which makes $S$ equal to 4K double precision words, or 32KB.

Our lower bound for the minimum number of cache misses that must be
suffered during a stencil computation holds for any cache, including 
fully associative caches.
The upper bound shows that a particular number of cache misses can be
achieved by choosing a special sequence of computations.
A \textit{cache miss} is defined as a request for a word of data that
is not present in the cache at the time of the request.
A \textit{cache load} is defined as an explicit request for a word of
data for which no explicit request has been made previously 
(a \textit{cold load}), or whose residence in the cache has expired 
because of a cache load of another word of data into the exact same
location in the cache (a \textit{replacement load}).
The definitions of cold and replacement loads match those of cold
and replacement cache misses, respectively \cite{MARTONOSI}, and
if $w$ equals 1 they completely coincide.

If a piece of code features $\phi$ cache misses and $\mu$ cache
loads, it can easily be shown that $\mu \leq w\phi$.
For a code with good spatial locality we typically have 
$\mu \approx w \phi$.
As can be shown by a simple example, 
no bound of the form $\phi \leq c \mu$ ($c$ constant) can be derived
for arbitrary code segments, but if the code implements a non-redundant
stencil operation, we have $\phi \leq |K| \mu$, where $|K|$ is the
total number of points within the stencil.
This is shown as follows.
Let the stencil operation be written as $q({\mathbf{x}}) = K u({\mathbf{x}})$,
with ${\mathbf{x}} \in \Omega$.
Here $\Omega$ is the (not necessarily contiguous) point set on which array
$q$ is evaluated.
Let $\overline{\Omega}$ be the $K$-extension of $\Omega$, which is the point
set on which $u$ must be defined in order to compute $q$ at all points of
$\Omega$.
The total distinct number of elements of $u$ used is $|\overline{\Omega}|$.
The number of cache misses $\phi$ does not exceed
the total number of accesses to array $u$ (may included repeated accesses
to the same element), which equals $|K| |\Omega|$, so 
$|\overline{\Omega}| \leq |K| |\Omega|$.
Consequently, we have the following interval inequality:
$|K|^{-1} \leq \frac{\mu}{\phi} \leq w$, which can be used to bound the
number of cache misses in terms of the number of cache loads.

\section{A lower bound for cache loads for local operators}\label{sec:LOWER}
In this section we consider the following problem:
for a given $d$-dimensional structured grid and a local stencil 
operator $K$, how many cache loads have to be incurred in order
to compute $q = Ku$, where $q$ and $u$ are two arrays defined
on the grid.
We will provide a lower bound $\mu$ which asserts that, regardless
of the order the grid points are visited for the computation of $q$,
at least $\mu$ cache loads have to occur.
In the next section we provide a \textit{cache fitting} algorithm for
the computation of $q$ whose number of cache loads closely
approaches the lower bound.

We use the following terminology to describe the operator $K$.
The vectors ${\mathbf{k}}_1$,\dots ,${\mathbf{k}}_s$ defined such
that $q({\mathbf{x}})$, the value of $q$ at the grid point identified
by the vector ${\mathbf{x}}$, is a function of the values
$u({\mathbf{x}}+{\mathbf{k}}_1)$,\dots ,$u({\mathbf{x}}+{\mathbf{k}}_s)$,
are called \textit{stencil vectors}.
Locality of $K$ means that the stencil vectors are contained in a cube
$\{{\mathbf{x}} | \, |x_i|\leq r, i=1,\dots ,d \}$ ($r$ is called the
\textit{radius} of $K$, and $2r\!+\!1$ its \textit{diameter}).
In this section we assume that $K$ contains only the star stencil
(i.e.\ the $\{{\mathbf{0}},{\mathbf{e}}_1,\dots ,{\mathbf{e}}_d,
-{\mathbf{e}}_1,\dots ,-{\mathbf{e}}_d\}$ stencil).
A lower bound for cache loads for the star stencil will give us a
lower bound for any stencil containing it.

Let $q$ be computed in the $K$-interior $R$ of a rectangular region
(a \textit{grid}) $G$.
We assume that computation of $q$ is performed in a pointwise fashion, 
that is, at any grid point the value of $q$ is computed completely before 
computation of the value of $q$ at another point is started.
In order to compute the value of $q$ at a grid point ${\mathbf{x}}$,
the values of $u$ in neighbor points of ${\mathbf{x}}$ must be loaded
into the cache (a point ${\mathbf{y}}$ is a neighbor of ${\mathbf{x}}$
if ${\mathbf{y\!-\!x}}$ is a stencil vector of $K$).
If ${\mathbf{x}}$ is a neighbor of ${\mathbf{y}}$ and $u({\mathbf{y}})$
has been loaded in cache to compute $q({\mathbf{z}})$ but is dropped from
the cache before $q({\mathbf{x}})$ is computed, then $u({\mathbf{y}})$
must be reloaded, resulting in a replacement load associated with 
${\mathbf{x}}$.

\begin{figure}[htb]
\centerline{\epsfig{file=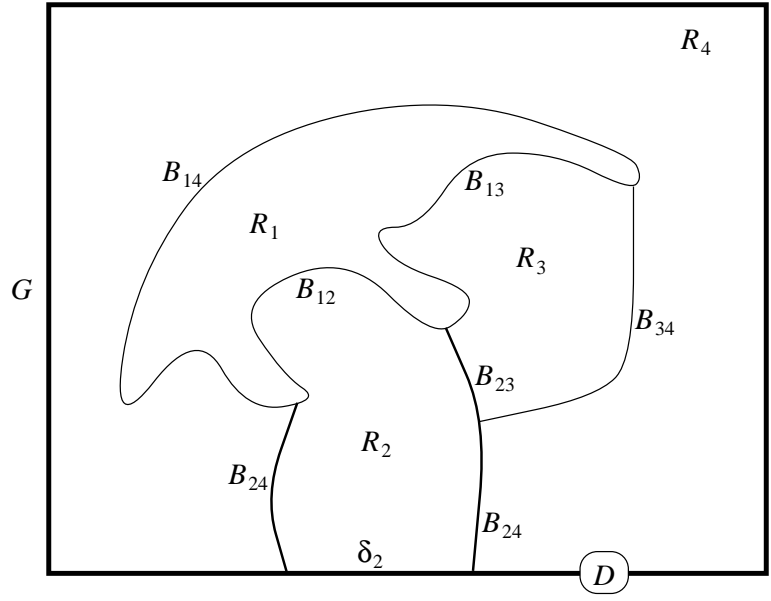,width=4in}}
\caption{\label{fig:ELEPHANT}The boundaries $B_{ij}$ of already computed
values of $q$ in a sequence of regions $R_i$.
Reloading of some values of $u$ on the boundary of $R_3$ results in at least
$\max (|B_{31}| + |B_{32}|-S,0)$ cache loads.}
\end{figure}

To estimate the number of elements, $\rho$, of array $u$ that must be
replaced, we choose a partition of $R$ into a disjoint union of grid
regions $R_i$, with $R = \cup_{i=1}^k R_i$, in such a way that $q$
is computed in all points of $R_i$ before it is computed at any point
of $R_{i+1}$, see Figure \ref{fig:ELEPHANT}.
Let $B_{ij}$ be the set of points in $R_j$ which are neighbors of
$R_i$.
Since the star stencil is symmetrical, the $B_{ij}$ are neighbor points of
$B_{ji}$.
Because any point of $B_{ij}$ can have at most $2d$ neighbors in $B_{ji}$,
we have the following inequalities:
\begin{equation} \label{eq:INTERVAL}
\frac{1}{2d} |B_{ij}| \leq |B_{ji}| \leq 2d |B_{ij}| \, .
\end{equation}

For computation of $q$ in $R_i$ we have to replace at least $\rho_i$
values of $u$, where $\rho_i$ equals $\max \left(\sum_{j=1}^{i-1} |B_{ij}|-S,0\right)$.
The total number of replaced values in the course of computing $q$ on the entire
grid will be at least $\rho$, where $\rho$ equals $B-kS$, and $B$ equals
$\sum_{i=1}^k \sum_{j=1}^{i-1} |B_{ij}|$.
Summing all the terms $|B_{ij}|$, taking into account Equation
(\ref{eq:INTERVAL}) and the fact that $|B_{ii}| = 0$, we get:
\begin{equation}
B \geq \frac{1}{4d} \sum_{i=1}^k \sum_{j=1}^{k} |B_{ij}| \, .
\end{equation}

Let $\delta R_i$ be the exterior boundary of $R_i$, that is, all points
neighbor to $R_i$ not belonging to $R_i$, and let $\delta_i$ be the subset
of the grid boundary $D$ having neighbors only in $R_i$.
Here, $D$ is defined as $G\setminus R$.
Obviously, $\delta_i \cap \delta_j = \emptyset$ if $i \neq j$, and
$\sum_{j=1}^k | B_{ij}| \geq |\delta R_i| - |\delta_i|$ .

Now we choose the $R_i$ such that $|\delta R_i| = \sigma \geq 8dS$, where
$\sigma$ is specified below, and let $\nu$ equal $\max |R_i|$.
Consequently, we have $k \geq |R|/\nu \mydef V/\nu$, and
\begin{equation}
\rho \geq \frac{1}{4d} \sum_{i=1}^k (|\delta R_i| - |\delta_i|) - kS =
k\left( \frac{\sigma}{4d}-S\right) - \frac{1}{4d} \sum_{i=1}^k |\delta_i| \geq
\frac{V}{\nu} S - \frac{1}{4d} |D| \, .
\end{equation}
We subsequently choose $\sigma$ in such a way that 
\begin{equation}
\sigma = |\delta O(d,t)| = \sum_{k=1}^d 2^k \binom{d}{k} \binom{t}{k-1} \geq 8dS 
\end{equation}
for some $t$, where $O(d,t)$ is the standard $d$-dimensional octahedron of radius $t$
(see Appendix A).
It follows from Equation \ref{eq:RADIUSLINK}, Appendix A, that $t$ can be chosen
in such a way that $\sigma$ is less than $8d(2d+1)S$.
Now the value of $\nu$ can be estimated using the isoperimetric property of
the octahedron (see again Appendix A), namely: $\nu \leq |O(d,t)|$.
Hence, we find
\begin{equation}
\frac{S}{\nu} \geq \frac{\sigma}{8d(2d+1)\nu} \geq
\frac{|\delta O(d,t)|}{8d(2d+1) |O(d,t)|} \geq c_d S^{-\frac{1}{d-1}} 
\end{equation}
where $c_d$ equals $1/(d(2d+1)2^{d+2})$.
This gives the following lower bound:
\begin{equation}
\rho \geq \frac{V}{8d(2d+1)} \frac{|\delta O(d,t)|}{|O(d,t)|} - \frac{1}{4d}|D| \geq
V c_d S^{-\frac{1}{d-1}} - \frac{1}{4d}|D| \, .
\end{equation}
We also have $V + |D| = |G|$ and $|D| \leq 2d |G|/l$, where $l$ is the smallest
size of the grid.
This gives the final lower bound $\mu$ for the total number of elements of $u$ to be
read into the cache:
\begin{equation} \label{eq:LOWER}
\mu \geq V+\rho \geq V\left(1+c_d S^{-\frac{1}{d-1}} -\frac{1}{2l}\right) \geq
    |G|\left(1 - \frac{2d+1}{l} + (1-\frac{2d}{l}) c_d S^{-\frac{1}{d-1}} \right) \, .
\end{equation}

In general, assuming that the cache associativity $a$ is larger than the diameter
of the operator $K$, the order of this lower bound can not be improved, as shows the 
following example (remember that our lower bound is valid for a cache with any 
associativity, including a fully associative cache).
Let the spatial extents of a two-dimensional grid be $n_1$ and $n_2$,
respectively, with $n_1$ equal to $kS$ and $n_2$ arbitrary,
and perform calculations of the star stencil (i.e.\ $r=1$) in the following order:

\begin{alltt}
   do   i = 0, k*a-1
     do   j = 2, \(n\sb{2}\)-1
       do   i1 = max(2,1+i*(S/a)), min(\(n\sb{1}\)-1,(i+1)*(S/a))
         q(i1,j) = u(i1,j) + \(\cdots\)
       end do
     end do
   end do
\end{alltt}

Since $n_1$ equals $kS$, all values of $q$ and $u$ having the same
value of the second index are mapped into the same cache location within a set.
Since $a$ exceeds $2r+1$, none of the values required for the computation of
$q$ will be replaced in the cache, except those at a distance $r$ around the
line defined by \texttt{i1} = \texttt{i*S/a}.
The total number of elements of $u$ read into the cache for execution of
this loop nest will therefore be $n_1 n_2 + (n_2-2)2r(ka-1)-4$, which
equals $n_1 n_2(1-2/n_1+2a(1-2/n_2)/S)$.
Similar examples in higher dimensions show that the order of our lower bound
(Equation \ref{eq:LOWER}) can not be improved.

\section{An upper bound for cache loads for local operators. 
         Cache fitting algorithm}\label{sec:UPPER}
In order to obtain an upper bound we present a \textit{cache fitting} algorithm
which has a small number of replacements.
We find a set $P$ of cache conflict-free indices of $u$ and calculate $Ku$
at the points of $P$.
Then we tile the index space of $u$ with $P$ to minimize the total number of
replacements.
For the analysis we assume an cache associativity of one, which is the worst
case for replacement loads.

\begin{figure}[htb]
\centerline{\epsfig{file=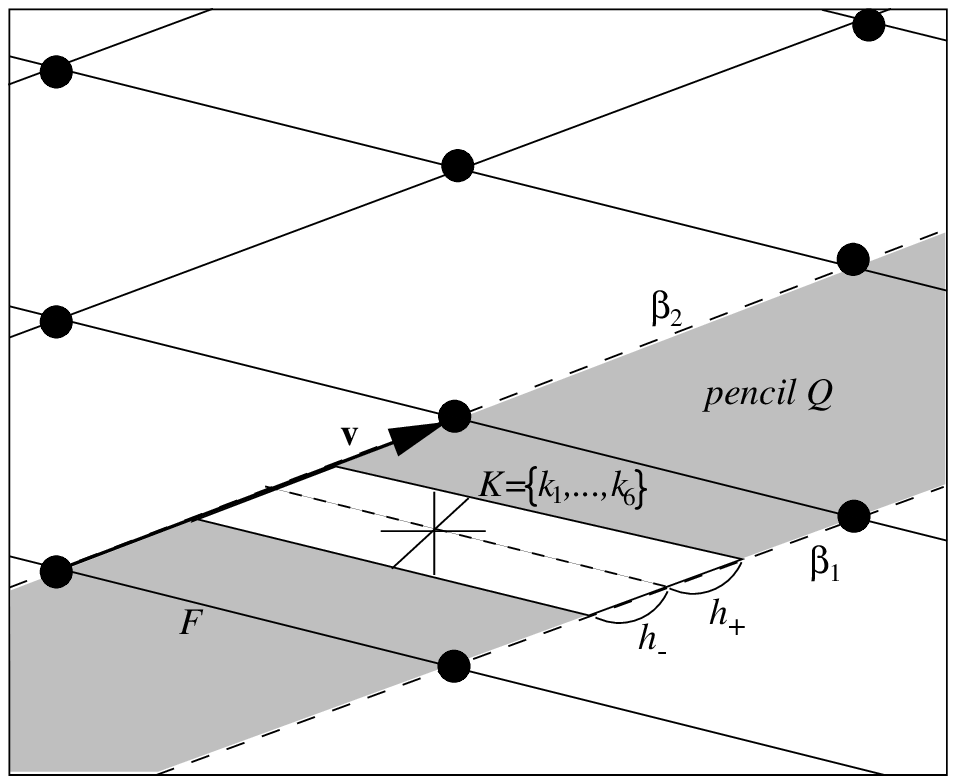,width=4in}}
\caption{\label{fig:LATTICE}\textbf{The Interference Lattice.}
Cache fitting set $F+k{\mathbf{w}}$, $k \in Z$, sweeps across pencil $Q$ in
the direction of ${\mathbf v}$.
Only values of $u$ at points at a distance $r$ or less from the pencil
boundaries $\beta_1$ and $\beta_2$ will be replaced in the cache when $K$ is
evaluated inside of $Q$.}
\end{figure}

Let $L$ be a set in the index space of $u$ having the same image in cache as
the index $(0,\dots,0)$, Figure \ref{fig:LATTICE}.
$L$ is a lattice in the sense that there is a generating set $\{{\mathbf b}_i\}$,
$i=1,\dots,d$, such that $L$ is the set of grid points 
$\{(0,\dots,0) + \sum_{i=1}^d x_i {\mathbf b}_i\, |\, x_i \in Z \}$.
We call this the \textit{interference lattice} of $u$.
It can be defined as the set of all vectors $(i_1,\dots,i_d)$ such that
\begin{equation}\label{eq:MISSEQ}
i_1 + n_1 i_2 + n_1 n_2 i_3 + \cdots + n_1 \cdots n_{d-1} i_d \equiv 0 \bmod S \, .
\end{equation}
In \cite{MARTONOSI} this lattice is defined as the set of solutions to the
cache miss equation.

Let $P$ be a fundamental parallelepiped of $L$\footnote{A fundamental parallelepiped
of a lattice $L$ is a set of points 
$\{ \sum_{i=1}^d x_i {\mathbf b}_i \, | \, 0 \leq x_i < 1\}$ for any basis
$\{ {\mathbf b}_i\}$ of $L$.}.
For future reference we note that $\vol(P) = \det L = S$.
The second equality follows form the fact that $L$ has a basis
$\{ {\mathbf v}_i\}$ of the form:
\begin{equation}\label{eq:BASIS}
{\mathbf v}_1 = S {\mathbf e}_1, \quad
{\mathbf v}_i = -m_i {\mathbf e}_1 + {\mathbf e}_i\, , 2\leq i \leq d \, ,
m_{i+1} = \prod_{j=1}^i n_j \, .
\end{equation}
Obviously, the vectors $\nu_i$ satisfy Equation \ref{eq:MISSEQ}.
Conversely, any vector satisfying Equation \ref{eq:MISSEQ} can be represented
as a linear combination of ${\mathbf v}_1,\dots,{\mathbf v}_d$, with coefficients
$x_k = i_k$ for $k=2,\dots,d$, and
$x_1 = (i_1 + m_2 i_2 + \dots + m_d i_d)S^{-1}$.
$x_1$ is an integer number, since $i_1 + m_2 i_2 + \dots + m_d i_d$ is divisible
by $S$ according to Equation \ref{eq:MISSEQ}.
Since ${\mathbf v}_1,\dots,{\mathbf v}_d$ are linearly independent vectors,
they form a basis of the lattice.

Let $F$ be a face of $P$ (see Figure \ref{fig:LATTICE}), and let ${\mathbf v}$ be
a basis vector of $L$ such that 
$P=\{ {\mathbf f}+x{\mathbf v} \,|\, {\mathbf f} \in F, 0 \leq x <1 \}$.
Then shifts $F+(k/g) {\mathbf v}$, $k=\dots,-1,0,1,\dots$ contain all integer points
of a pencil $Q$, with 
$Q=\{ {\mathbf f}+x{\mathbf v} \,|\, 
      {\mathbf f} \in F, x \hbox{\textit{ is any number}} \}$
for an appropriate value of $g$\footnote{Let $I$ be a fundamental parallelepiped of
the integer lattice in the subspace $Y$ generated by $F$, and let ${\mathbf e}$ be
an integer vector such that ${\mathbf e}$ and the basis of $I$ generate $Z^d$.
Obviously, $g$ must be chosen in such a way that $\vol((1/g){\mathbf v},I) =
\vol({\mathbf e},I) = 1$.
Hence, $g = \vol({\mathbf v},I) = (1/|F|) \vol({\mathbf v},F) = \vol(P)/|F|$,
where $|F|$ is the index of the lattice $L \cap Y$ in the integer lattice of Y.
}.
The values of $q$ at the points of $Q$ can be computed without replacing reusable
values of $u$ except at a distance of $r$ or less from the boundary of $Q$.
Let $h_1,\dots, h_s$ be the signed projections along $F$ of the stencil vectors
of $K$ onto ${\mathbf v}$, and let $h_+$ and $h_-$ be the maximum and the 
minimum of the projections, respectively.
We assume also that $|h_+-h_-|/g < |{\mathbf v}| a$, meaning that the extent
of $P$ in the direction of ${\mathbf v}$ is big enough to allow to compute
$q$ on $F$ without replacements.
It may be impossible to satisfy this condition when the shortest vector in $L$
is shorter than the diameter of $K$ divided by the cache associativity.
Lattices with short vectors are discussed in Section \ref{sec:RESULTS}.
The associated grids are called \textit{unfavorable grids}.

The \textit{Cache Fitting Algorithm} for computing $q$ is as follows
(see Figure \ref{fig:LATTICE});
here $K(R)$ is the set of points where $u$ must be available in order to compute
$q$ in all points of $R$ (i.e. the $K$-extension of $R$):

\begin{alltt}
   set \({\mathbf w}\) = (1/g)\({\mathbf v}\)
   do   Q = Qmin, Qmax
     determine face \(F\) inside pencil Q
     do   \(k\) = kmin, kmax
       load in cache all values of \(u\) inside \(K(F+k*{\mathbf{w}})\)
       compute \(q\) at \(F+k*{\mathbf{w}}\)
     end do
   end do
\end{alltt}

In this algorithm the parameters \texttt{Qmin}, \texttt{Qmax}, \texttt{kmin}
and \texttt{kmax} are determined such that the scanning face $F$ sweeps out the
entire grid.
Whenever a point is not contained in the grid, it is simply skipped in the
nest.

Since we defined the algorithm in such a way that the scanning face in the direction
of ${\mathbf v}$ with step $g$ passes through all integer points of $Q$, the
values of $q$ at all points inside $Q$ will be computed.

Replacements misses can occur only at points at a distance of $r$ or less from the
boundaries of the pencils.
For each of these points at most $s$ replacement need to take place, where $s$,
the size of the stencil, is defined by $s = |K| \leq (2r+1)^d$.
So the number of replacements will not exceed $r(2r+1)^d A$, where $A$ is the
total surface area of all pencils.

To minimize $A$ we choose $P$ so that $Q$ has a good surface to volume ratio.
Let $P$ be the fundamental parallelepiped of a reduced basis of $L$.
A basis ${\mathbf b}_1,\dots,{\mathbf b}_d$ of a $d$-dimensional lattice $L$
is called reduced if 
\begin{equation}\label{eq:REDUCED}
\prod_i ||{\mathbf b}_i || \leq c_d \det L \, ,
\end{equation}
 where $c_d$ is a constant which depends only on $d$%
\footnote{Every lattice has a reduced basis.
There is a polynomial algorithm to find a reduced basis with a constant
$c_d = 2^{d(d-1)/4}$ \cite[Ch.\ 6.2]{SCHRIJVER}.
}.
Let ${\mathbf b}_1$ be the shortest vector of the basis, and let the eccentricity
$e$ of the basis be defined by $e=\max(||{\mathbf b}_i||/||{\mathbf b}_1||)$.
If we define $\partial P$ as the surface of $P$, we can derive an estimation for 
the surface-to-volume ratio of $P$:
\begin{equation}\!\!\!\!\!\!\! \label{eq:SURFACETOVOLUME}
\frac{|\partial P|}{\det L} \leq 
\frac{2 \sum_j \prod_{i \neq j} ||{\mathbf b}_i||}{\det L} \leq
2 c_d \sum_i\frac{1}{||{\mathbf b}_i||} \leq 
c^{\prime}_d \frac{1}{||{\mathbf b}_1||} \leq
e c^{\prime}_d \left(\prod_i ||{\mathbf b}_i||\right)^{-\frac{1}{d}} \!\!\leq
e c^{\prime}_d S^{-\frac{1}{d}} \, ,
\end{equation}
where we twice used the Hadamard inequality:
$\prod_i ||{\mathbf b}_i|| \geq \det L$,
and the abovementioned fact that $L=S$.
The constant $c^{\prime}_d$ is defined by $c^{\prime}_d = 2d c_d$.

Since $A$ does not exceed the surface area of all fundamental parallelepipeds
covering the grid, the total number of these parallelepipeds (which equals
$|G|/\det L$) gives us: $A \leq |\partial P| |G| / \det L$, so that the
total number of replacements $\rho$ can be bounded by:
$\rho \leq r(2r+1)^d|\partial P| |G| /\det L$.
This, combined with Equation \ref{eq:SURFACETOVOLUME}, gives an upper bound
for the total number of elements to be loaded into the cache in the
cache fitting algorithm:
\begin{equation} \label{eq:UPPER}
\mu \leq |G| + \rho \leq |G| \left( 1+e c^{\prime\prime}_d S^{-\frac{1}{d}}\right) \, ,
\end{equation}
where $c^{\prime\prime}_d$ is defined by $c^{\prime\prime}_d = r(2r+1)^d c^{\prime}_d$.

Note that if the shortest vector in the interference lattice has length
$(S/f)^{1/d}$ for some constant $f$ it follows that $e<f c_d$.
To show this, we sort the basis vectors in Equation \ref{eq:REDUCED} in ascending
order.
Then it follows that $\frac{S}{f}^{\frac{d-1}{d}} ||{\mathbf b}_d|| \leq c_d S$,
and hence
$e = \frac{||{\mathbf b}_d||}{||{\mathbf b}_1||} \leq f c_d$.

In Appendix B we show that there are grids whose interference lattices feature
$f$'s that are independent of $S$ (provided that $S$ is a prime power, which is
true in most practical cases).
For these lattices the relative gap between the upper bound (Equation \ref{eq:UPPER}) 
and the lower bound (Equation \ref{eq:LOWER}) of the previous section goes to zero as $S$
increases.
When the cache associativity exceeds the diameter of $K$, this gap can be closed.
In that case a parallelepiped, built on a reduced basis of the interference 
lattice of the array indices with $x_d=0$, can be swept in the $d^{th}$
coordinate direction, similar to the example at the end of Section \ref{sec:LOWER}.
In general, the cache fitting algorithm gives full cache utilization, in contrast
to the algorithm for finding grid-aligned parallelepipeds devoid of interference
lattice points, as proposed in \cite{MARTONOSI}.
See Table 2, \cite{MARTONOSI}, where the sizes of blocks without self interference
are approximately 20\% smaller than $S$.

\section{Lower and upper bounds for multiple RHS arrays}
In this section we consider the case where there multiple arrays involved
in the computation of $q$.
Let $p$ be the number of arrays (we call these the \textit{RHS arrays}), all
having the same sizes, and let the stencil of each RHS array include the
star stencil.
This means, in particular, that for each boundary point of any region $R_i$
(see Figure \ref{fig:ELEPHANT}) values of $p$ RHS arrays are necessary%
\footnote{As in Section \ref{sec:LOWER}, we assume that computation of $q$ is 
performed in a pointwise fashion.
In this case elements of all RHS arrays have to be loaded into cache
simultaneously, reducing the cache size effectively by a factor of $p$.
Non-pointwise computations may be performed if the operator $K$ is
separable, in the sense that it can be written as
$K(u_1,u_2,\dots,u_p) = K_1(u_1,K_2(u_2,\dots K_p(u_p)\dots)$.
In the case of separability of $K$ the stencil operation can be split into
a succession of independent operations, each involving an intermediate
value of $q$ and one RHS array.
This would not require to load all $p$ RHS arrays in cache at each point.
Instead, it would suffice to write intermediate values of $q$ into main
memory, and then load them back into cache for completion of the
computations.
This results in a larger effective cache size, but more data to be
loaded, so splitting the operation need not improve the total number of loads.}
for computation of $q$ in $R_i$.
Hence, we have to replace at least $\rho_i$ values, with
$\rho_i = \max(p(\sum_{j<i}|B_{ij}|)-S,0)$ values of RHS arrays.
Now we can repeat the arguments of Section \ref{sec:LOWER}, with
$|V|$ and $|G|$ replaced by $p|V|$ and $p|G|$, respectively, and
$S$ replaced by $\lceil S/p \rceil$, to obtain the following lower bound for the
number of cache loads for stencil computations with $p$ RHS arrays:
\begin{eqnarray}
\mu \geq p |V| + \rho &\geq&
p |V| \left(1+c_d \left\lceil \frac{S}{p}\right\rceil^{-\frac{1}{d-1}} - 
\frac{1}{2l}\right)  \nonumber\\
&\geq&
p |G| \left(1-\frac{2d-1}{l} + \left(1-\frac{2d}{l}\right)
c_d \left\lceil \frac{S}{p}\right\rceil^{-\frac{1}{d-1}}\right) \, .
\end{eqnarray}

\begin{figure}[htb]\begin{center}
\epsfig{file=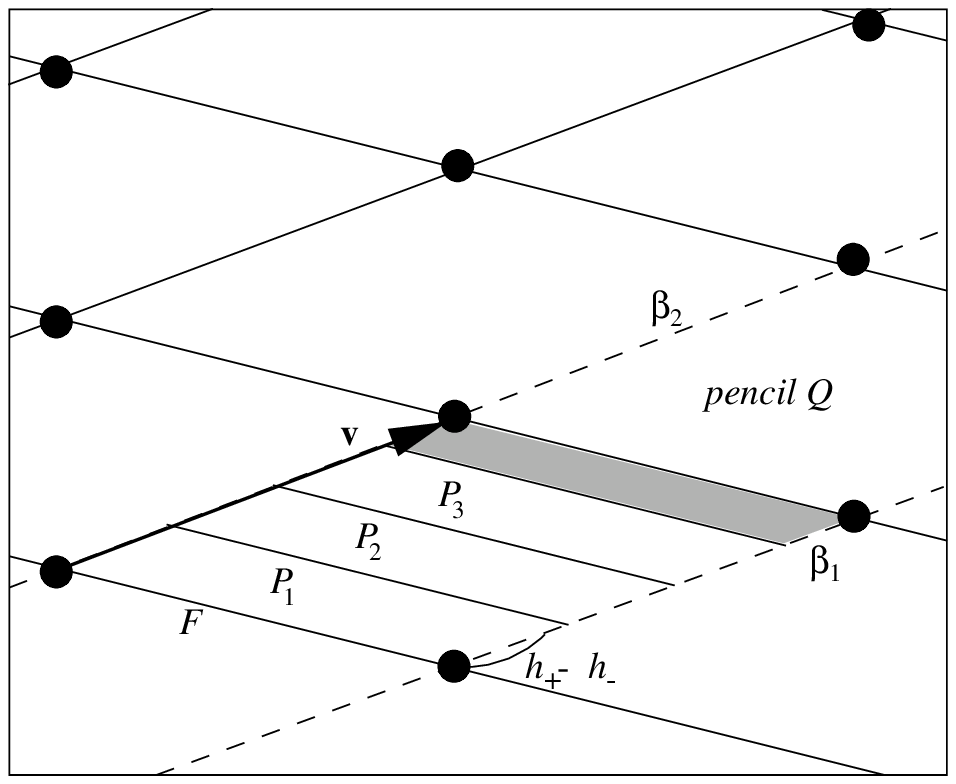,width=3.5truein}
\caption{\label{fig:MULTILATTICE}Tiling of a fundamental parallelepiped
of a reduced basis of the lattice $L$.
We assume that $|h_+ - h_-| \leq a |{\mathbf{v}}/p|$ ($a$ is the cache associativity).
The tiling effectively reduces the size of the parallelepiped by a factor of
at most $2p$ (since $x/p \geq \lfloor x/p \rfloor \geq x/(2p)$), and increases the 
cost of a replacement in the cache per point of the boundary of the pencil by at most 
a factor of $p$, since elements of all $p$ RHS arrays will be replaced at the same 
time.}
\end{center}
\end{figure}

In order to obtain an upper bound for cache loads for calculations with
$p$ RHS arrays, we assume that we are free to choose relative array offsets.
Our upper bound is valid on the assumption that the stencil diameter divided 
by the cache associativity is smaller than the length of the longest lattice 
basis vector divided by $p$.
Consider a stripwise tiling of the fundamental parallelepiped $P$ for the
lattice $L$, see Figure \ref{fig:MULTILATTICE}.
Each tile $P_i$ has the same size and shape.
The size is determined by considering the longest edge vector ${\mathbf{v}}$
in the fundamental parallelepiped and dividing it into $p$ equal pieces of size
$[(S/|F|)/p] ||{\mathbf{v}}||$, so that each tile contains $|F| [(S/|F|)/p]$
integer points, where $|F|$ is the number of integer points in the face.
The remainder part of the tiling is indicated by the shaded area.
The reason why the longest edge vector is selected for subdivision is as
follows.
Since we use a reduced basis, the smallest angle between ${\mathbf{v}}$ and $F$
is bounded from below, so the parallelepiped is always close to orthogonal.
Therefore, subdividing the longest edge leads to tiles with the largest
inscribed sphere, and thus the largest difference stencil fitting inside
the tile.

Let $\{P_i\}$ be the parallelepipeds of the tiling, and let $s_i$ be the
address offset of $P_i$ relative to $P_1$ (corresponding to the same RHS array).
We assign one parallelepiped to each RHS array and choose starting addresses
of the arrays, $\hbox{\textrm{addr}}_i$, in such a way that images of tiles
$P_i$ in the cache do not overlap: 
$\hbox{\textrm{addr}}_i = \hbox{\textrm{addr}}_1 + m_iS+s_i$, where
$m_1=s_1=0$, and $m_i=m_{i-1}+\lceil\frac{|V|-s_i+s_{i-1}}{S}\rceil, i=2,\dots,p$.
Sweeping through the pencil by units of tile $P_1$ in the direction of
${\mathbf{v}}$ we can compute $Ku$ without any cache conflicts, except on the
boundary of the pencils.
The number of replacement loads of this algorithm can be estimated similarly to 
the number of replacement loads of a single-array algorithm, taking into account
that for calculation of a value $u$ at any point values of all $p$ RHS arrays
in the neighbor points may have to be in cache, thus reducing the effective cache
size to $[S/p]$:
\begin{equation}
\mu \leq p|G| + \rho \leq
p|G| \left(1+e c^{\prime\prime}_d \left[\frac{S}{p}\right]^{-\frac{1}{d}}\right) 
\end{equation}
where $c^{\prime\prime}_d$ is a constant which depends only on $d$, and $e$ is
the eccentricity of $L$.

\section{Unfavorable array sizes}\label{sec:RESULTS}
\begin{figure}[htb]\begin{center}
\centerline{\epsfig{file=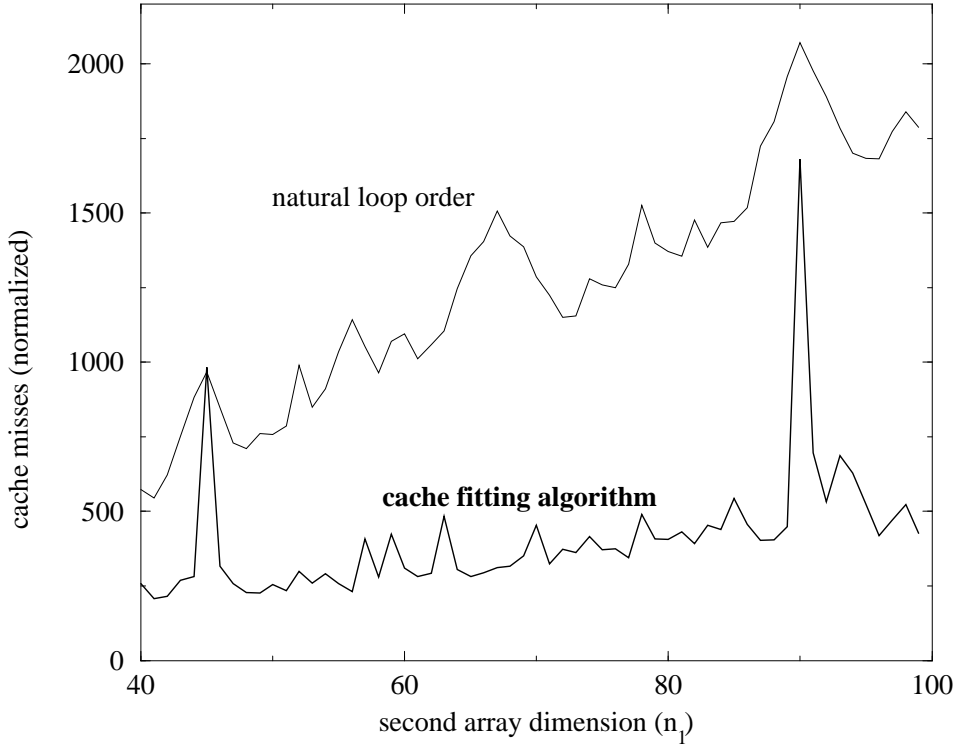,width=5in}}
\caption{\label{fig:MISSES}
Plot of measured cache misses for $40\leq n_1 < 100$, $n_2=91$ for 13-point
star stencil.
The top line corresponds to the naturally ordered nest, optimized by the
SGI Fortran compiler.
The bottom line corresponds to our cache fitting algorithm.
A typical ratio between the two is 3.5.
The large fluctuations correspond to grids with short lattice vectors
($n_1=45$ and $n_1=90$ yield shortest vectors $(1,0,1)$ and $(2,0,1)$,
respectively).
The fluctuations of cache misses of the cache fitting algorithm for such
grids can be so big that their cache misses become more numerous than for
the compiler-optimized nest.
}
\end{center}
\end{figure}

We have implemented our cache fitting algorithm and compared its actually
measured number of cache misses with those of the compiler-optimized code 
for the corresponding naturally ordered loop nest on a MIPS R10000 processor
(SGI Origin 2000).
For comparison we chose a second order difference operator
(the common 13-point star stencil) an a test set including three-dimensional
grids of sizes $40\leq n_1 < 100$, $n_2=91$, and $n_3=100$ (the value of the
second dimension was chosen to show a typical picture;
that of the third dimension is irrelevant).
A plot of measured cache misses for both codes is shown in Figure
\ref{fig:MISSES}.
The program was compiled with options ``\texttt{-O3 -LNO:prefetch=0},''
using the MIPSpro f77 compiler, version 7.3.1.1m.
The prefetch flag disables the prefetching compiler optimization.
Without this option the number of cache misses increases significantly, because
the compiler does aggressive prefetching to try to reduce execution time.

The upper bounds for the cache misses from the previous sections would suggest
that the number of replacement cache misses will increase in the cases where
the interference lattice has a very short vector.
Very short means that the length is smaller than the diameter of the operator
divided by the cache associativity.
In this case the self interference would increase significantly.
This result suggests how to pad arrays to improve cache performance:
the padding should be organized in such a way that the shortest vector in the
lattice is not too short, though short enough to minimize the number of pencils
(large index of scanning face $F$).
The sweeping  is organized such that pencils are as wide as possible
(i.e.\ the smallest total number of pencils), while avoiding---in the case
of multiple RHS arrays---tiles that are thinner than the diameter of the 
stencil operator divided by the cache associativity.

\begin{figure}[htb]\begin{center}
\epsfig{file=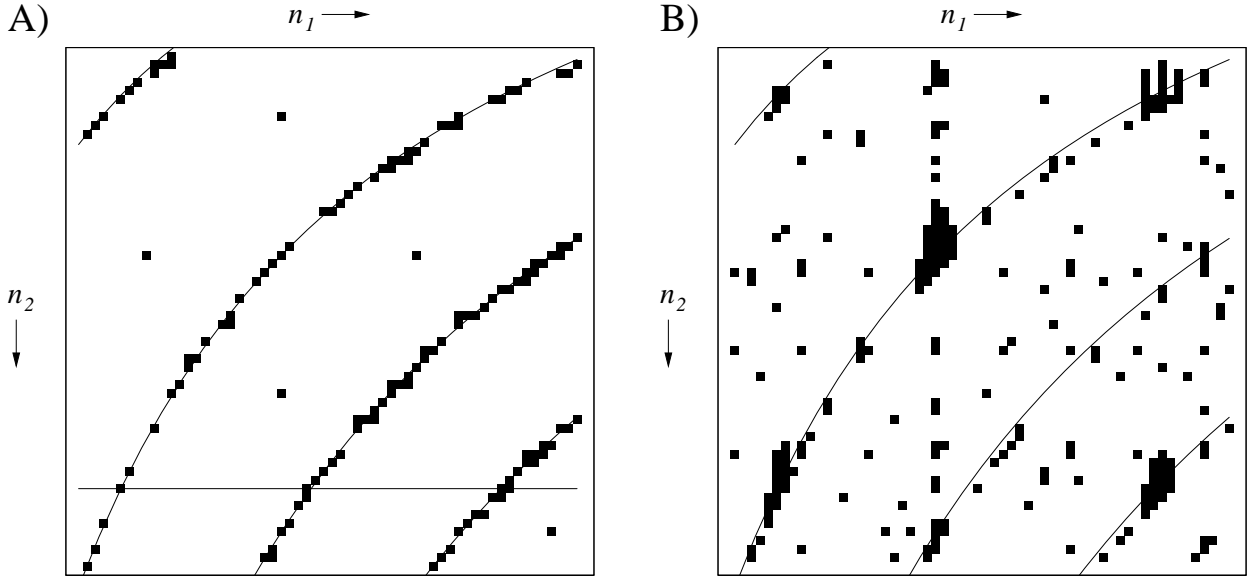,width=6.5truein}
\caption{\label{fig:CORRELATE}Plot A shows measured fluctuations of cache 
misses (above 15\% of the upper bound).
Plot B shows the interference lattices with short (less than 8 in the $L_1$ norm)
vectors.
Array sizes are $40\leq n_1, n_2 <100$.
The plots can be fitted well by hyperbolae defined by $n_1 n_2 = \frac{1}{2}kS,
k=1,2,3,4$, meaning that arrays with unfavorable size are those whose $z$-slices 
are (close to) multiples of half the cache size.
The horizontal line in Plot A shows the position of the graph from Figure
\ref{fig:MISSES}.}
\end{center}
\end{figure}

To demonstrate these unfavorable grids we again choose the second order stencil
and force computations in the nest to follow the natural order\footnote{%
This forcing is accomplished by introducing a dependence through
a Fortran subroutine that performs a circular shift of its arguments}.
Figure \ref{fig:CORRELATE}a shows the correlation between spikes in the number 
of cache misses and the presence of a very short vector in the lattice.
We call these lattices unfavorable for cache utilization.
Arrays having such lattices should be avoided on the target machine.
When the shortest vector of the interference lattice is shorter than the
diameter of the operator, the number of cache misses sharply increases.
The application developer should avoid such unfavorable array sizes, and
compilers should avoid the sizes using appropriate padding of array dimensions.
Note that similar unfavorable cache effects have been mentioned in \cite{BAILEY}.

\section{Conclusions and future work}
We have demonstrated tight lower and upper bounds for cache misses for
calculations of an explicit operator $K$ on a structured grid.
Our lower bound is valid in the general case of fully associative caches,
and is based on a discrete isoperimetric theorem.
Our upper bound is based on a cache fitting algorithm which uses the fundamental
parallelepiped of a special basis of the interference lattice to fit the data
in the cache.
The upper bound assumes that the shortest vector in the interference lattice is
not too short.
We have shown that there are grids whose interference lattices have this property.
We have also shown that the presence of a very short vector in the lattice
correlates with fluctuations of actual cache misses for calculation of a second
order explicit operator on three-dimensional grids.
The fluctuations occur on grids with unfavorable sizes, i.e.\ on those whose
product of the first two dimensions is (close to) a multiple of half the cache size.

Our results can be extended straightforwardly to implicit stencil computations 
(i.e.\ those of the form $q \leftarrow K(q)$) when the problem has a one-dimensional
data dependence.
Such a data dependence exists if computations of $q$ at grid points can take place in 
an arbitrary order, except that there is a single index $i$ for which 
$q(x_1,\dots,i,\dots,x_d)$ must be evaluated before $q(x_1,\dots,i+\alpha,\dots,x_d)$
can be calculated (the constant $\alpha$ is either +1 or -1).
Clearly, the lower bound is not affected by the implicitness of $K$.
The previously derived upper bound can still be achieved by prescribing the proper
visit order of points within each parallelepiped, of the scanning face direction
within each pencil (positive or negative sweep direction), and of the visit order
of subsequent pencils.  
This is always possible for a one-dimensional data dependency.

Our results can also be extended to arrays that store more than one word per grid
point (tensor arrays).
The lower bound of Section \ref{sec:LOWER} for operations with multiple right hand
sides immediately applies to tensor arrays.
The upper bound of that section also applies, provided the tensor components can 
be stored as independent subarrays.

In a future study we plan to extend the results of this paper to more general
implicit operators, to operators on unstructured grids, and to tensor arrays
with restricted storage models.
We intend to study more closely the dependence of cache misses on the size
of the operator's stencil.
We also plan to enhance the presented results by taking into account a secondary
cache and TLB, and to formulate bounds for cache misses more directly than
through the determination of cache loads.
\section*{Appendix A: The simplex and the octahedron}
In this section we list some basic facts on the number of integer points in the
octahedron and simplex.
The standard octahedron is defined as:
\begin{equation}
O(d,t) = \left\{{\mathbf{x}} \in Z^d \, \mid \, \sum_{i=1}^d |x_i| \leq t\right\} 
\end{equation}
and the standard simplex as:
\begin{equation}
S(d,t) = \left\{{\mathbf{x}} \in Z^d \, \mid \, 0\leq x_1,\dots,x_d \, ,
\sum_{i=1}^d |x_i| \leq t\right\} \, .
\end{equation}
If we consider sections of the octahedron by planes $x_1=k$, $k=-t,\dots,t$, then
for the number of integer points in the octahedron we get the following 
recurrence relation:
\begin{equation}
|O(d,t)| = |O(d-1,t)| + 2\sum_{k=0}^{t-1} |O(d-1,k)| \, .
\end{equation}
This relation can be used to prove that
\begin{equation}\label{eq:OCTAGON}
|O(d,t)| = \sum_{k=0}^d 2^k \binom{d}{k} \binom{t}{k}
\end{equation}
and that
\begin{equation}
|\delta O(d,t-1)| = |O(d,t) - O(d,t-1)| = 
 \sum_{k=1}^d 2^k \binom{d}{k}\binom{t-1}{k-1} \, .
\end{equation}
Also, the relation 
\begin{equation}
|\delta O(d,t)| = |\delta O(d,t-1)| + |\delta O(d-1,t)| + |\delta O(d-1,t-1)|
\end{equation}
shows that
\begin{equation}\label{eq:RADIUSLINK}
|\delta O(d,t)| \leq (2d+1) |\delta O(d,t-1)| \, .
\end{equation}
For the number of integer points in the simplex we have the following recurrence
relation:
\begin{equation}
|S(d,t)| = |S(d-1,t)| + |S(d,t-1)| \, .
\end{equation}
This can be used to prove that cf.\ \cite{KNUTH}, Table 169, see also \cite{WANG},
Section 5:
\begin{equation}\label{eq:SIMPLEX}
|S(d,t)| = \sum_{k=1}^d \binom{d}{k}\binom{t}{k} = \binom{d+t}{d} \, .
\end{equation}
From Equations \ref{eq:OCTAGON} and \ref{eq:RADIUSLINK} it follows that
$|O(d,t)| \leq 2^d |S(d,t)|$.
Also, since $\delta O(d,t-1)$ contains at least two nonoverlapping simplices
$S(d-1,t)$ and can be covered by $2^d$ such simplices, we see that
\begin{equation}
2|S(d-1,t)| \leq |\delta O(d,t-1)| \leq 2^d |S(d-1,t)| \, , \,\,d\geq 2 \, .
\end{equation}
Hence, if $|S(d-1,t)|$ equals $S$, we have for $d\geq 2$:
\begin{equation}\label{eq:ISOPERIMETRIC}
\frac{|\delta O(d,t)|}{|O(d,t)|} \geq
\frac{|\delta O(d-1,t)|}{|O(d,t)|} \geq
\frac{2 |S(d-1,t)|}{2^d |S(d,t)|} = 
2^{-d+1}\left(1+\frac{t}{d}\right)^{-1} \geq 2^{-d+1} S^{-\frac{1}{d-1}} \, ,
\end{equation}
since from Equation \ref{eq:SIMPLEX} it follows that if
$|S(d-1,t)|$ does not exceed $S$, then $1+t/d$ does not exceed $S^{1/(d-1)}$.

The \textit{isoperimetric inequality} \cite{WANG}, Theorem 2, asserts that the
size of the boundary of a subset $R$ in $Z^d$ is at least as big as the size of the
standard sphere that contains $|R|$ points\footnote{%
The standard sphere, defined in \cite{WANG}, is the integer point set of minimal
surface area for any given number of interior points.}.
It is easy to see that any standard sphere is sandwiched between two standard
octahedrons whose radii differ by one.
Since the octahedron has the largest volume for a given fixed-size boundary,
Inequality \ref{eq:ISOPERIMETRIC} is true for any lattice body with a boundary of 
size $S$.

\section*{Appendix B: The existence of grids with favorable lattices}
In order to prove that for every cache of size $S=p^n$, where $p$ is a 
prime number, there are grids with interference lattices whose shortest
vector has a length $l$ greater than $(S/f)^{1/d}$, with $f$ independent
of $S$, we show:
\begin{itemize}
\item[a.] For every dimensionality $d$ there exists a lattice $L$ of the
          same dimension whose basis has the form given in Equation \ref{eq:BASIS}
          (Section \ref{sec:UPPER}), and whose shortest vector is sufficiently
          long, and
\item[b.] a grid can be constructed that has $L$ as its interference lattice.
\end{itemize}

\noindent\textit{Corollary:}
Since grids with dimensions $n_i+k_i S$, $i=1,\dots ,d$ have the same interference
lattice for any non-negative integers $k_i$, any grid can be embedded in a favorable
larger grid.

\vspace{.4cm}
\noindent Proof:

\begin{itemize}
\setlength{\itemindent}{.5cm}
\item[Step a:] Let a lattice $L$ have a basis of the form of Equation \ref{eq:BASIS}.
Any lattice vector ${\mathbf{x}}$, which includes all basis vectors of $L$, with 
$L^\infty$ norm at most $l$ must be a solution of the following system of 
inequalities:
\begin{equation}\label{eq:SYSTEM}
\left. \begin{array}{c} |x_i| \leq l \, , i=2,\dots ,d \\
                        |Sx_1 + m_2 x_2 + \dots + m_d x_d| \leq l
       \end{array} \right\} \, .
\end{equation}
Existence of a solution to this system is equivalent to that of the system
\begin{equation}
\left. \begin{array}{c} 
|x_i| \leq l \, , i=2,\dots ,d \\
\left\|\dfrac{m_2}{S} x_2 + \dots + \dfrac{m_d}{S} x_d\right\|
\leq \dfrac{l}{S}
       \end{array} \right\} 
\end{equation}
where $\|z\|$ is the distance from $z$ to the nearest integer number.
Theorem VIII, Ch.\ 1 \cite{CASSELS} states that there are real numbers
$\mu_2,\dots, \mu_d$, and a constant $c_d^{\prime\prime\prime}$ depending
only on $d$, such that 
\begin{equation}
\|\mu_2 x_2 + \dots + \mu_d x_d \| \geq \dfrac{c_d^{\prime\prime\prime}}{l^{d-1}}
\end{equation}
for all nonzero ${\mathbf{x}}$ satisfying $|x_i| \leq l \, , i=2,\dots ,d$.

If we choose the nonzero integers $m_i$ in such a way that 
$|m_i-S\mu_i| \leq 2$ for $i=2,\dots,d$, then  we get
\begin{equation}
\left\|\dfrac{m_2}{S} x_2 + \dots + \dfrac{m_d}{S} x_d\right\| \geq
\dfrac{c_d^{\prime\prime\prime}}{l^{d-1}} - (d-1)\dfrac{l}{S} =
\left(\dfrac{S}{l^d} c_d^{\prime\prime\prime} -(d-1)\right)\dfrac{l}{S}
\end{equation}
which shows that Equation \ref{eq:SYSTEM} has no integer solutions if 
$l < (c_d^{\prime\prime\prime}/(Sd))^{1/d}$.
Hence, $f$ in Section \ref{sec:UPPER} can be chosen as: $f=d/c_d^{\prime\prime\prime}$,
and the lattice with the basis given by Equation \ref{eq:BASIS} has a reduced
basis with eccentricity depending only on $d$.

\item[Step b:]
In order to find a grid whose interference lattice is $L$, we first sort the $m_i$
in order of increasing gcd$(m_i,S)$.
Since we assume that $S=p^n$, where $p$ is prime, we know that gcd$(m_i,S)$
divides gcd$(m_{i+1},S)$, and the appropriate grid dimensions $n_i$ can be found
directly by solving the congruencies $(n_i m_i - m_{i+1}) \bmod S = 0$.
\end{itemize}

\section*{Acknowledgements.}
We are grateful to professor Leonid Khachijan for some help with the theoretical
part of the paper, and to Jerry Yan and Henry Jin for discussions on the practical
aspects of the paper.
This work is partially supported by the HPCC/CAS NASA program, and was executed
under NAS task order A61812D. 

\end{document}